# Adaptive Submodular Optimization under Matroid Constraints

Daniel Golovin and Andreas Krause

*California Institute of Technology, Pasadena, CA 91125, USA*

**Abstract**

Many important problems in discrete optimization require maximization of a monotonic submodular function subject to matroid constraints. For these problems, a simple greedy algorithm is guaranteed to obtain near-optimal solutions. In this article, we extend this classic result to a general class of adaptive optimization problems under partial observability, where each choice can depend on observations resulting from past choices. Specifically, we prove that a natural adaptive greedy algorithm provides a $1/(p+1)$ approximation for the problem of maximizing an adaptive monotone submodular function subject to $p$ matroid constraints, and more generally over arbitrary $p$-independence systems. We illustrate the usefulness of our result on a complex adaptive match-making application.

## 1. Introduction

Submodular functions play an important role in discrete optimization. Many important problems, such as facility location [1], coverage [2], influence maximization [3], and experimental design [4] can be reduced to constrained maximization of a submodular set function. Submodularity, informally, is an intuitive notion of diminishing returns, which states that adding an element to a small set helps more than adding that same element to a larger (super-)set. While maximizing submodular functions in general is NP-hard, a celebrated result of Nemhauser et al. [1] shows that a simple greedy algorithm is guaranteed to find a near-optimal solution to the problem of maximizing a submodular function subject to cardinality constraints; the value of the greedy solution obtains at least a constant fraction of $(1 - 1/e)$ of the optimal value. Nemhauser and Wolsey [5] furthermore show that, in general, obtaining better approximation guarantees requires evaluating the objective function on an exponential number of sets. Beyond cardinality constraints, the greedy algorithm also obtains guarantees for maximizing a monotone submodular function subject to $p$ matroid constraints [6], and, more generally, $p$-independence systems [7]. In both cases, the greedy algorithm achieves at least a constant fraction of $1/(p+1)$ of the optimal value.

Submodular optimization provides a unified framework for many important non–adaptive optimization problems. In many practical optimization problems, however, one needs to *adaptively* make a sequence of decisions, taking into account observations about the outcomes of past decisions. Often these outcomes are uncertain, and one may only know a probability distribution over them. Finding optimal policies for decision making in such partially observable stochastic optimization problems is notoriously intractable (see, e.g., [8]). The classical notion of submodular set functions does not allow one to handle such adaptive optimization problems. In recent work [9, 10], we introduced the concept of *adaptive submodularity*, a natural generalization of submodular set functions to adaptive policies, and prove that if a partially observable stochastic optimization problem satisfies this property, a simple adaptive greedy algorithm is guaranteed to obtain near-optimal solutions in the case of cardinality constraints. The concept has several useful applications, including active learning, machine diagnosis, adaptive viral marketing, and sensor placement [10, 11].

In this article, we prove that the adaptive greedy algorithm is guaranteed to obtain a $1/(p+1)$-approximation for the more general problem of maximizing an adaptive submodular function over $p$-independence systems (and therefore subject to $p$ matroid constraints). Our results generalize those obtained by Asadpour et al. [12] for optimizing a particular instance of adaptive submodular functions to a single matroid constraint to arbitrary adaptive monotone submodular functions and to arbitrary $p$-independence systems. We illustrate the usefulness of this result on adaptive match–making problems such as online dating. We show that the constraint system in this application is a 2-independence system, and thus the adaptive greedy algorithm provides a $1/3$-approximation.

## 2. Background

We first review background on submodular optimization, as well as the recently discovered adaptive submodularity property.

### 2.1. Submodular optimization over $p$-independence systems

Let $E$ be a finite set of items that we may select among. Consider an objective function $f : 2^E \to \mathbb{R}$. For each $e \in E$ let $\Delta(e \mid A) := f(A \cup \{e\}) - f(A)$ denote the *marginal benefit* of item $e$ w.r.t. set $A$. We call $f$ *monotone* iff $\Delta(e \mid A) \geq 0$ for all $e \in E$ and $A \subseteq E$. We call $f$ *submodular*, iff for all

$A \subset B \subseteq E$ and $e \in E \setminus B$ it holds that $\Delta(e \mid A) \geq \Delta(e \mid B)$, i.e., adding $e$ to a set $A$ provides larger marginal benefit than adding it to set $B$. Many *non-adaptive* problems in discrete optimization can be reduced to finding a set $A$ that maximizes a monotone submodular function $f$ subject to some constraints.

One natural class of constraints that we wish to consider are *matroid constraints*, which require the solution to be an *independent set* of a matroid:

**Definition 1 (Matroid).** *A matroid is a pair $(E, \mathcal{I})$ with a nonempty collection $\mathcal{I} \subseteq 2^E$ of* independent sets *such that*
- *For all $A \subset B \subseteq E$, if $B \in \mathcal{I}$ then $A \in \mathcal{I}$.*
- *For all $A, B \in \mathcal{I}$ with $|B| > |A|$, there exists $e \in B \setminus A$ such that $A \cup \{e\} \in \mathcal{I}$.*

Matroids were developed as an abstraction of families of independent objects, such as sets of linearly independent vectors in a vector space. One important example is the *uniform matroid*, where sets are independent iff they contain at most $k$ items for some fixed $k$, i.e., we wish to pick a set $A$ that maximizes $f$ over all sets of size at most $k$. Other important examples include *partition matroids* and *graphical matroids*.

A yet more general class of constraints are $p$-*independence system constraints*, which require the solution to be an independent set of a $p$-independence system:

**Definition 2 ($p$-independence system).** *For a parameter $p \in \mathbb{N}$, a $p$-independence system is a pair $(E, \mathcal{I})$ with a nonempty collection $\mathcal{I} \subseteq 2^E$ of* independent sets *such that*
- *For all $A \subset B \subseteq E$, if $B \in \mathcal{I}$ then $A \in \mathcal{I}$.*
- *For $C \subseteq E$, let $\mathcal{B}(C)$ be the set of maximal elements of $\mathcal{I}$ which are subsets of $C$. That is, $\mathcal{B}(C) := \{B \subseteq C : \forall e \in C \setminus B, B \cup \{e\} \notin \mathcal{I}\}$. Then for all nonempty $C \subseteq E$, we have $\max_{B \in \mathcal{B}(C)} |B| \leq p \cdot \min_{B \in \mathcal{B}(C)} |B|$.*

One important special case of a $p$-independence systems is the intersection of $p$ matroids: $(E, \cap_{i=1}^p \mathcal{I}_i)$ where $(E, \mathcal{I}_i)$ is a matroid for all $i$. Calinescu et al. [7] discuss other special cases, such as $p$-circuit-bounded families considered in [13] and $p$-extendible families defined in [14].

While maximizing submodular functions is hard even for a single uniform matroid constraint, a celebrated result of Fisher et al. [6] is that a simple greedy algorithm obtains a $1/(p+1)$ approximation for the problem of maximizing a monotone submodular function subject to the intersection of $p$ matroids. Hereby, the greedy algorithm (under constraint $\mathcal{I}$) starts with the empty set $A = \emptyset$, and iteratively adds an item

$$e^* \in \arg\max_{e \,:\, A \cup \{e\} \in \mathcal{I}} \Delta(e \mid A). \qquad (1)$$

to $A$ until $A$ is a maximal set in $\mathcal{I}$.

In some applications, implementing the greedy rule (1) may by itself require solving an NP-hard optimization problem (e.g., selecting tours for information–gathering robots [15]). Fortunately, the proof given in [6] can be extended to show that for any $\alpha \in [0, 1]$ an $\alpha$-*approximate greedy algorithm* obtains an approximation guarantee of $\frac{\alpha}{p+\alpha}$. Here, an $\alpha$-approximate greedy algorithm is one that selects an $\alpha$–approximation to the best greedy element in each step, i.e., an item $e^*$ such that $A \cup \{e^*\} \in \mathcal{I}$ and $\Delta(e \mid A) \geq \alpha \cdot \max_{e \,:\, A \cup \{e\} \in \mathcal{I}} \Delta(e \mid A)$, until $A$ is maximal in $\mathcal{I}$. Furthermore, this result holds also for the more general case of a $p$-independence system constraint. Extending the proof to an $\alpha$-approximate greedy algorithm was done by Goundan and Schulz in the case of the intersection of $p$ matroids [16], and by Calinescu et al. for the more general $p$-independence system constraint [7].

## 2.2. Adaptive optimization

We are interested in *adaptive* optimization problems, where our choice of items can depend on observations about items chosen previously. We will now formalize the class of adaptive optimization problems that we consider.

Each item $e \in E$ is in a particular (initially unknown) state $\Phi(e) \in O$ from a set $O$ of possible states. Hereby, $\Phi : E \to O$ is a (random) *realization* of the ground set, indicating which state each item is in. We take a Bayesian approach and assume that there is a (known) probability distribution $\mathbb{P}[\Phi]$ over realizations. We will consider problems where we sequentially pick an item $e \in E$, get to see its state $\Phi(e)$, pick the next item, get to see its state, and so on. After each pick, our observations so far can be represented as a *partial realization* $\Psi$, a function from some subset of $E$ (i.e., the set of items that we already picked) to their states. For notational convenience, we sometimes represent $\Psi$ as a relation, so that $\Psi \subseteq E \times O$ equals $\{(e, o) : \Psi(e) = o\}$. We use the notation $\text{dom}(\Psi) = \{e : \exists o.(e, o) \in \Psi\}$ to refer to the domain of $\Psi$ (i.e., the set of items observed in $\Psi$). A partial realization $\Psi$ is *consistent* with a realization $\Phi$ if they are equal everywhere in the domain of $\Psi$. In this case we write $\Phi \sim \Psi$. If $\Psi$ and $\Psi'$ are both consistent with some $\Phi$, and $\text{dom}(\Psi) \subseteq \text{dom}(\Psi')$, we say $\Psi$ is a *subrealization* of $\Psi'$. Equivalently, $\Psi$ is a subrealization of $\Psi'$ if and only if $\Psi \subseteq \Psi'$.

We encode our adaptive strategy for picking items as a *policy* $\pi$, which is a function from a set of partial realizations to $E$, specifying which item to pick next under a particular set of observations. If $\Psi \notin \text{dom}(\pi)$, the policy terminates (stops picking items) upon observation of $\Psi$. Technically, we require that the domain of $\pi$ is closed under subrealizations. That is, if $\Psi' \in \text{dom}(\pi)$ and $\Psi$ is a subrealization of $\Psi'$ then $\Psi \in \text{dom}(\pi)$. We use the notation $E(\pi, \Phi)$ to refer to the set of items selected by $\pi$ under realization $\Phi$.

We wish to maximize, subject to some constraints, a utility function $f : 2^E \times O^E \to \mathbb{R}_{\geq 0}$ that depends on which items we pick and which state each item is in. Based on this notation, the expected utility of a policy $\pi$ is $f_{\text{avg}}(\pi) := \mathbb{E}[f(E(\pi, \Phi), \Phi)]$ where the expectation is taken with respect to $\mathbb{P}[\Phi]$.

In this paper, we address the problem of optimization subject to a $p$-independence system. Thus, our goal is to find a policy $\pi^*$ such that

$$\pi^* \in \arg\max_\pi f_{\text{avg}}(\pi) \text{ subject to } E(\pi, \Phi) \in \mathcal{I} \text{ for all } \Phi, \quad (2)$$

where $\mathcal{I}$ is a collection of independent sets defined by a $p$-independence system.

*2.3. Adaptive submodularity*

In general, Problem (2) cannot be approximated to within a factor of $\mathcal{O}(|E|^{1-\epsilon})$ for any constant $\epsilon > 0$ under the reasonable complexity–theoretic assumption that $PH \neq \Sigma_2^P$ [10], even if $f$ is modular in its first argument for all realizations, i.e., even if there exist $w_{e,\Phi} \geq 0$ such that $f(A, \Phi) = \sum_{e \in A} w_{e,\Phi}$. However, there is a natural adaptive generalization of monotone submodular functions, such that many classic results on the optimization of monotone submodular functions generalize to the adaptive realm [10]. These adaptive generalizations are defined in terms of the *conditional expected marginal benefits* of items, that is, the expected marginal benefit of an item, conditioned on the current partial realization.

**Definition 3 (Conditional Expected Marginal Benefit).**
*Given a partial realization $\Psi$ and an item $e$, the* conditional expected marginal benefit *of $e$ conditioned on having observed $\Psi$, denoted $\Delta(e \mid \Psi)$, is*

$$\Delta(e|\Psi) := \mathbb{E}\left[f(\mathrm{dom}(\Psi) \cup \{e\}, \Phi) - f(\mathrm{dom}(\Psi), \Phi) | \Phi \sim \Psi\right] \quad (3)$$

*where the expectation is taken with respect to $\mathbb{P}[\Phi]$.*

We are now ready to introduce our generalizations of monotonicity and submodularity to the adaptive setting:

**Definition 4 (Adaptive Monotonicity [10]).** *A function $f : 2^E \times O^E \to \mathbb{R}_{\geq 0}$ is* adaptive monotone *with respect to distribution $\mathbb{P}[\Phi]$ if the conditional expected marginal benefit of any item is nonnegative, i.e., for all $\Psi$ with $\mathbb{P}[\Psi] > 0$ and all $e \in E$ we have*

$$\Delta(e|\Psi) \geq 0. \quad (4)$$

**Definition 5 (Adaptive Submodularity [10]).** *A function $f : 2^E \times O^E \to \mathbb{R}_{\geq 0}$ is* adaptive submodular *with respect to distribution $\mathbb{P}[\Phi]$ if the conditional expected marginal benefit of any fixed item does not increase as more items are selected and their states are observed. Formally, $f$ is adaptive submodular w.r.t. $\mathbb{P}[\Phi]$ if for all $\Psi$ and $\Psi'$ such that $\Psi$ is a subrealization of $\Psi'$ (i.e., $\Psi \subseteq \Psi'$), and for all $e \in E$, we have*

$$\Delta(e|\Psi') \leq \Delta(e|\Psi). \quad (5)$$

Note the similarity with the submodularity condition, which we wrote as $\Delta(e \mid B) \leq \Delta(e \mid A)$ for all $A \subset B \subseteq E$ and $e \in E \setminus B$ specifically to highlight the connection. When $\mathbb{P}[\cdot]$ is deterministic, so that only one realization can occur, the adaptive monotonicity and adaptive submodularity reduce to the classic monotonicity and submodularity conditions.

The power of these definitions is that certain results for monotone submodular maximization generalize from (non-adaptive) sets to (adaptive) policies [10]. In this paper, we will show that this is true even for the general case of maximizing a monotone submodular function under an arbitrary $p$-independence system constraint.

## 3. Adaptive Optimization Over $p$-Independence Systems

In this section, we analyze a natural adaptive generalization of the greedy algorithm (1) and its $\alpha$-approximate versions to the adaptive realm. We call our adaptive generalization of an $\alpha$-approximate algorithm an $\alpha$*-approximate greedy policy*.

**Definition 6 ($\alpha$-approximate greedy policy, for constraint $\mathcal{I}$).**
*A policy $\pi$ is an $\alpha$-approximate greedy policy[1] for constraint $\mathcal{I}$ if, upon observing partial realization $\Psi$, $\pi$ selects an item $e$ such that $\mathrm{dom}(\Psi) \cup \{e\} \in \mathcal{I}$ (if such an $e$ exists) and such that*

$$\Delta(e \mid \Psi) \geq \alpha \cdot \left( \max_{e' : \mathrm{dom}(\Psi) \cup \{e'\} \in \mathcal{I}} \Delta(e' \mid \Psi) \right). \quad (6)$$

*Otherwise, if no such an item $e$ exists, $\pi$ terminates.*

We are now ready to prove our main result: The adaptive generalization of the non-adaptive $\frac{\alpha}{p+\alpha}$ approximation guarantee of an $\alpha$-approximate greedy algorithm in the case of a $p$–independence system constraint.

**Theorem 7.** *Fix an adaptive monotone submodular function $f : 2^E \times O^E \to \mathbb{R}_{\geq 0}$ and a $p$-independence system $(E, \mathcal{I})$. Fix a policy $\pi$ which is $\alpha$-approximate greedy with respect to $f$ for constraint $\mathcal{I}$. Then $\pi$ yields an $\frac{\alpha}{p+\alpha}$ approximation, meaning*

$$f_{avg}(\pi) \geq \left( \frac{\alpha}{p+\alpha} \right) \max_{\text{feasible } \pi^*} f_{avg}(\pi^*)$$

*where $\pi^*$ is feasible iff $E(\pi^*, \Phi) \in \mathcal{I}$ for all $\Phi$.*

PROOF. Our proof is an adaptive generalization of the elegant proof of Calinescu et al. [7] for the non–adaptive case, which itself is based on a scheme of Jenkyns [13]. We will not reproduce that proof here, but instead point out the features of it that we need. Consider the non–adaptive case, and fix any maximal, feasible set of elements $S = \{e_1, e_2, \ldots, e_k\}$. Formally, $S \in \mathcal{I}$ and $S \cup \{e\} \notin \mathcal{I}$ for all $e \in E \setminus S$. Let $S_i := \{e_1, e_2, \ldots, e_i\}$ for all $i$. Let $S^* \in \mathcal{I}$ be any feasible set. As a key step in their analysis, Calinescu et al. prove the existence of a sequence of sets $S_1^*, S_2^*, \ldots, S_k^*$ whose nonempty members partition $S^*$, such that for all $i$ and $e \in S_i^*$, $S_{i-1} \cup \{e\} \in \mathcal{I}$, and $|S_i^*| \leq p$ for all $i$. We will call $S_1^*, \ldots, S_k^*$ a *decomposition of $S^*$ with respect to the sequence* $(e_1, e_2, \ldots, e_k)$.

Now consider the adaptive case. Fix $\alpha$-approximate policy $\pi$ and any feasible policy $\pi^*$. Let $\Psi_{i,\Phi}$ be the partial realization observed by $\pi$ after selecting $\min\{i, k(\Phi)\}$ elements under true realization $\Phi$, where $k(\Phi) := |E(\pi, \Phi)|$. We consider the decomposition of $E(\pi^*, \Phi)$ with respect to the sequence of elements selected by $\pi$ under $\Phi$, in the order they are selected by $\pi$. Let $E_1^*(\Phi), \ldots, E_{k(\Phi)}^*(\Phi)$ be such a decomposition. Note that each $e \in E_{i+1}^*(\Phi)$ is feasible when the algorithm has observed $\Psi_{i,\Phi}$, in the sense that $\mathrm{dom}(\Psi_{i,\Phi}) \cup \{e\} \in \mathcal{I}$. Hence, any $\alpha$-approximate policy $\pi$ must obtain conditional marginal expected benefit $\Delta\left(\pi(\Psi_{i,\Phi})\,|\,\Psi_{i,\Phi}\right) \geq \alpha \cdot \Delta\left(e\,|\,\Psi_{i,\Phi}\right)$ for all $e \in E_{i+1}^*(\Phi)$. Since $|E_{i+1}^*(\Phi)| \leq p$, as all parts of the decomposition promised by Calinescu et al. must be, we have

$$\Delta\left(\pi(\Psi_{i,\Phi})\,|\,\Psi_{i,\Phi}\right) \geq \frac{\alpha}{p} \sum_{e \in E_{i+1}^*(\Phi)} \Delta\left(e\,|\,\Psi_{i,\Phi}\right) \quad (7)$$

---
[1]In [10], $\alpha$-approximate greedy policies are defined for $\alpha \geq 1$, whereas here we define them for $\alpha \leq 1$ (as the inverse of the factor in [10]) to maintain uniformity with previous work on matroid constraints and their generalizations.

Note we can write $f_{\text{avg}}(\pi)$ in the form

$$f_{\text{avg}}(\pi) = \sum_{\Psi:\Psi \in \text{dom}(\pi)} \mathbb{P}\left[\Phi \sim \Psi\right] \cdot \Delta\left(\pi(\Psi)\,|\,\Psi\right) \quad (8)$$

where $\Phi$ is the true realization. Intuitively, if we consider the decision tree associated with $\pi$ — where items are selected at nodes and the branches are determined by the selected item's state — then the previous equation is obtained by using linearity of expectation to decompose the expected benefit of $\pi$ into a weighted sum of conditional expected marginal benefits of each node in the tree, where the weight of a node is the probability it is reached. We can then rearrange these terms, as

$$f_{\text{avg}}(\pi) = \sum_{\Psi:\Psi \in \text{dom}(\pi)} \sum_{\Phi:\Phi \sim \Psi} \mathbb{P}\left[\Phi\right] \cdot \Delta\left(\pi(\Psi)\,|\,\Psi\right) \quad (9)$$

$$= \sum_{\Phi} \mathbb{P}\left[\Phi\right] \sum_{\Psi:\Phi \sim \Psi} \Delta\left(\pi(\Psi)\,|\,\Psi\right) \quad (10)$$

where $\Delta\left(\pi(\Psi)\,|\,\Psi\right) = 0$ whenever $\Psi \notin \text{dom}(\pi)$. Rewriting Eq. (10) using the $\Psi_{i,\Phi}$ notation and they applying Eq. (7) yields

$$f_{\text{avg}}(\pi) = \sum_{\Phi} \mathbb{P}\left[\Phi\right] \sum_{i=0}^{k(\Phi)-1} \Delta\left(\pi\left(\Psi_{i,\Phi}\right)\,|\,\Psi_{i,\Phi}\right) \quad (11)$$

$$\geq \frac{\alpha}{p} \sum_{\Phi} \mathbb{P}\left[\Phi\right] \sum_{i=0}^{k(\Phi)-1} \sum_{e \in E_{i+1}^*(\Phi)} \Delta\left(e\,|\,\Psi_{i,\Phi}\right) \quad (12)$$

Next we require the definition of *policy concatenation* [10]: Given two policies $\pi_1$ and $\pi_2$ define $\pi_1@\pi_2$ as the policy obtained by running $\pi_1$ to completion, and then running policy $\pi_2$ as if from a fresh start, ignoring the information gathered during the running of $\pi_1$. Hence under realization $\Phi$ the policy $\pi_1@\pi_2$ selects the union of items selected by $\pi_1$ and $\pi_2$ under $\Phi$. Thus, $E(\pi_1@\pi_2, \Phi) = E(\pi_1, \Phi) \cup E(\pi_2, \Phi)$.

Now for each $(e, \Phi)$ such that $e \in E(\pi^*, \Phi)$ define $\Psi_{e,\Phi}$ as the partial realization observed by $\pi@\pi^*$ immediately before selecting $e$. We will assume without loss of generality that each element is selected by $\pi@\pi^*$ at most once. (If this is not the case, we may for purposes of analysis replace the groundset $E$ with $E \times \mathbb{N}$, order the instances in which $\pi@\pi^*$ selects $e$ arbitrarily, and replace the $j^{\text{th}}$ instance of $e$ with $(e, j)$. The objective function then simply treats all copies $\{(e, j) : j \in \mathbb{N}\}$ the same, as do the realizations.) Then for all $e \in E_{i+1}^*(\Phi)$, $\Delta(e\,|\,\Psi_{i,\Phi}) \geq \Delta(e\,|\,\Psi_{e,\Phi})$ by adaptive submodularity, since $\Psi_{i,\Phi} \subseteq \Psi_{e,\Phi}$. Let $RHS_{12}$ denote the right-hand-side of Eq. (12). Combining the above fact with Eq. (12) yields

$$RHS_{12} \geq \frac{\alpha}{p} \sum_{\Phi} \mathbb{P}\left[\Phi\right] \sum_{i=0}^{k(\Phi)-1} \sum_{e \in E_{i+1}^*(\Phi)} \Delta\left(e\,|\,\Psi_{e,\Phi}\right) \quad (13)$$

$$= \frac{\alpha}{p} \left(f_{\text{avg}}\left(\pi@\pi^*\right) - f_{\text{avg}}\left(\pi\right)\right) \quad (14)$$

$$\geq \frac{\alpha}{p} \left(f_{\text{avg}}\left(\pi^*\right) - f_{\text{avg}}\left(\pi\right)\right) \quad (15)$$

To obtain Eq. (14), note that $\{E_{i+1}^*(\Phi) : 0 \leq i \leq k(\Phi)-1\}$ partitions $E(\pi^*, \Phi)$, and we can decompose $f_{\text{avg}}\left(\pi@\pi^*\right) - f_{\text{avg}}\left(\pi\right)$ in a manner similar to how we obtained Eq. (11). Intuitively, in Eq. (14) we are summing contributions over exactly those nodes in the decision tree corresponding to $\pi@\pi^*$ which are there due to $\pi^*$, and not over those there due to $\pi$. Lastly, Eq. (15) follows from the adaptive monotonicity of $f$, because it implies $f_{\text{avg}}(\pi@\pi^*) \geq f_{\text{avg}}(\pi^*)$. For a proof of this last fact, see Lemma 30 of [10].

Finally, combining Eq. (11) through Eq. (15) yields $f_{\text{avg}}(\pi) \geq \frac{\alpha}{p}\left(f_{\text{avg}}\left(\pi^*\right) - f_{\text{avg}}\left(\pi\right)\right)$ which may be rearranged to yield the claimed bound of $f_{\text{avg}}(\pi) \geq \left(\frac{\alpha}{p+\alpha}\right) f_{\text{avg}}(\pi^*)$. $\square$

## 4. Application to Adaptive Match-Making

In this section we consider an application of Theorem 7 to adaptive match-making problems such as online dating. We imagine we have an undirected graph $G = (V, E)$ where $V$ is the set of people using our match-making service, and $E$ is a set of feasible pairings of people, i.e., $\{u, v\} \in E$ if $v$ meets the requirements specified by $u$ and vice–versa. The service will recommend dates (corresponding to edges) to people, and after each date the participants will provide feedback $o$ on the quality of their experience. We suppose there is a known distribution over how well each date $\{u, v\}$ will turn out, based on the profiles of $u$ and $v$. We also suppose that each person $v$ has specified an upper bound $d(v)$ on the number of dates desired. Subject to this constraint, we seek to maximize a known monotone submodular function $\hat{f}$ of sets of (date, feedback) pairs. For example, we could define a *good* date as one where the participants wish to go on another date together, let the feedback be whether the date was good or not, and define $\hat{f}$ as the number of users who experienced at least one good date in the input set of (date, feedback) pairs. As another example for a feasible $\hat{f}$, the feedback for a date $\{u, v\}$ could be a pair of numeric scores corresponding to how satisfied $u$ and $v$ were with the date, respectively, and $\hat{f}$ could be the sum of *user scores*, where the score of a user $u$ is the maximum score that $u$ gave to any date. There are many other possibilities.

Hence, we seek to solve (2) where the items are edges, $\Phi(e)$ encodes the feedback from each date $e$, and $\mathcal{I}$ consists of all subsets of edges whose induced subgraphs satisfy the degree constraints that each $v$ has degree at most $d(v)$. Formally, $\mathcal{I} := \{A \subseteq E : \forall v \in V, \deg_A(v) \leq d(v)\}$ where $\deg_A(v) := |\{w : \{v, w\} \in A\}|$. Assuming the feedback from each date is an independent sample from its distribution, i.e., $\mathbb{P}[\Phi] = \prod_{e \in E} \mathbb{P}[\Phi(e)]$, then we have the following result.

**Theorem 8.** *Any $\alpha$-approximate greedy policy $\pi$ achieves $\frac{\alpha}{2+\alpha}$ as much reward in expectation as the optimal policy. In particular, the greedy policy achieves a $1/3$-approximation to the optimal policy.*

We spend the remainder of this section proving Theorem 8. We start by establishing that in our model the underlying objective, $f(A, \Phi) := \hat{f}(\{(e, \Phi(e)) : e \in A\})$, is adaptive monotone submodular. This follows from the following theorem, whose proof appears in [10].

**Theorem 9** (§6 of [10]). *Fix a monotone submodular function $\hat{f} : 2^{E \times O} \to \mathbb{R}_{\geq 0}$ and a prior $\mathbb{P}[\Phi]$ with independent outcomes, so that $\mathbb{P}[\Phi] = \prod_{e \in E} \mathbb{P}[\Phi(e)]$. Let $f(A, \Phi) := \hat{f}(\{(e, \Phi(e)) : e \in A\})$. Then $f$ is adaptive monotone submodular.*

Given $f$ is adaptive monotone submodular, to prove Theorem 8 it suffices to show that the constraints in the match-making problem can be modeled by a $p$-independence system with $p = 2$, since then we can apply Theorem 7 to complete the proof. We now proceed to do so.

**Theorem 10.** *For all $d : V \to \mathbb{N}$, $(E, \mathcal{I})$ is a 2-independence system.*

PROOF. If $G$ is bipartite, with $V = U \uplus W$ being the bipartition (so that $E \subseteq U \times W$), then this is easy to prove by establishing that $\mathcal{I}$ is the intersection of two matroids $(E, \mathcal{I}_U)$ and $(E, \mathcal{I}_W)$, where $\mathcal{I}_S := \{A \subseteq E : \forall v \in S, \deg_A(v) \leq d(v)\}$ for all $S \in \{U, W\}$.

We prove the case of general graphs using a more direct approach. Recall Definition 2, and fix any $C \subseteq E$. Fix $A$ and $B$ to be arbitrary maximal independent subsets of $C$.

For each $\{u, w\} \in B$, charge one unit to $\arg\min_{v \in \{u,w\}} \{d(v) - \deg_A(v)\}$. Break ties arbitrarily. Let $c(v)$ be the total resulting charge to $v$. We claim for all $v$,

$$c(v) \leq \deg_A(v) \qquad (16)$$

Suppose, by way of contradiction, that $c(v) > \deg_A(v)$. Then some $\{u, v\} \in B \setminus A$ is charged to $v$, and so $d(v) - \deg_A(v) \leq d(u) - \deg_A(u)$. Note $c(w) \leq \deg_B(w) \leq d(w)$ for all $w$. These facts together imply

$$1 \leq c(v) - \deg_A(v) \leq d(v) - \deg_A(v) \leq d(u) - \deg_A(u)$$

Hence $\deg_A(u) < d(u)$ and $\deg_A(v) < d(v)$, which implies $A \cup \{\{u, v\}\} \in \mathcal{I}$, contradicting the maximality of $A$.

Given Eq. (16), and the basic fact that $\sum_v \deg_E(v) = 2|E|$ in any graph $(V, E)$, we have

$$|B| = \sum_v c(v) \leq \sum_v \deg_A(v) = 2|A|$$

which completes the proof. □

## 5. Conclusions

Adaptive submodularity [10] provides an elegant framework for analyzing myopic strategies for certain adaptive optimization problems, in much the same way that submodularity does for certain non–adaptive optimization problems. In this article we have shown how the greedy policy achieves near–optimal performance under a large parameterized class of constraints, the $p$-independence system constraints, and shown how to apply this result in the context of a complex adaptive match–making application. We believe that this and related results in [10] will prove useful for several other applications involving adaptive optimization.